\documentclass{iaus}
\usepackage{graphicx}

\title[Surface temperature maps for II Peg during 1999-2002]
{Surface temperature maps for II Peg during 1999-2002}

\author[Lindborg, M. et al.]
{Lindborg, M.$^{1}$, Korpi, M.J.$^{1,4}$, Tuominen, I.$^1$, Hackman, T.$^1$, Ilyin, I$^2$, Piskunov, N.$^3$}

\affiliation{$^1$Observatory, PO BOX 14, FI-00014 University of Helsinki, Finland \\[\affilskip]
$^2$Astrophysikalisches Institut Potsdam, An der Sternwarte 16, 14482 Potsdam, Germany \\[\affilskip]
$^3$Department of Astronomy and Space Physics, Uppsala University,
SE-751 20, Uppsala, Sweden\\[\affilskip]
$^4$NORDITA, Roslagstullsbacken 23, SE-10691 Stockholm, Sweden}

\pubyear{2009}
\volume{264}  
\pagerange{119--126}
\jname{Solar and Stellar Variability Impact on Earth and Planets}
\editors{A.C. Editor, B.D. Editor \& C.E. Editor, eds.}
\begin{document}

\maketitle

\begin{abstract}
The active RS CVn star II Peg has been spectroscopically monitored for
almost 18 years with the SOFIN spectrograph at NOT, La Palma,
Spain. In this paper we present five new surface temperature maps of the
object for the years 1999 (two maps), 2001 (one map) and 2002 (two maps). 
\end{abstract}

\firstsection 
\section{Introduction}

II Peg is one of the very active RS CVn stars and it is the brightest X-ray star with 50pc of the Sun. RS CVn stars
are closely detached binaries where the more massive
component is a G-K giant or subgiant and the secondary a subgiant
or dwarf of spectral class G to M. Because of the low luminosity of
the secondary many RS CVn systems appear as single-line binaries
which are suitable for spectral analysis (Berdyugina et
al. 1998a). In close binaries the rapid rotation is maintained by
tidal forces due to the close companion.  Large amplitude brightness
variation of RS CVn stars imply the presence of enormous star spots
on their surfaces covering up to 50\% of the visible disk. Also
coronal X-ray and microwave emissions, strong flares in the optical,
UV, radio and X-ray are seen. Cool spots on the stellar surface
will locally alter the photospheric absorption lines and continuum
intensities.

Previous investigations on the temperature distribution over the
surface of II Peg include the study of Berdyugina et al. (1998b), who
presented surface temperature maps for 1992-1996, and Bergyugina et
al. (1999c) with surface maps for 1996-1999, both studies were based on
observations with the SOFIN-spectrograph at NOT. Gu et al. (2003)
presented surface images of II Peg for 1999-2001 based on observations
with the Coude echelle spectrograph at the Xinglong station of the
National Astronomical Observatories, China, but the spectral lines
used for inversions were different to that of SOFIN
observations. Photometric light curve variations of the object were
analysed by Berdyugina \& Tuominen (1998c), and by Rodon\`{o} et
al. (2000).  The results of Berdyugina et al. (1998a,b, 1999a,b,c) and
Berdyugina \& Tuominen (1998c) consistently show that there are two
active longitudes separated approximately by 180$^{\circ}$, migrating
in the orbital reference frame, and that a switch of activity level
occurs periodically with a period of 4.65 years. In the surface images
of Gu et al. (2003) the general spot pattern was quite similar, but
the drift with respect of the orbital frame was less obvious, and the
switch of the activity level appeared to occur earlier than predicted
by Berdyugina et al (1999c). Rodon\`{o} et al. (2000) found a much more complicated
spot pattern from their analysis of photometry: they report on the
existence of a longitudinally uniformly distributed component together
with three active longitudes, with complicated cyclic behavior.
Carroll et al. (2009) have also applied a Zeeman Doppler imaging
technique to derive the magnetic field configuration on the surface of
II Peg during 2007 using spectropolarimetric (Stokes I and V)
observations with SOFIN. Their maps show a very similar spot pattern
as found by Berdyugina et al. (1998b, 1999c); moreover, the radial
field direction is opposite on different active longitudes.

\section{Observations}

\begin{table}
  \begin{center}
    \caption{Summary of the observations.}
    \label{table1}
    \begin{tabular}{lccclccc}
    \hline
    HJD      &Phase &S/N  &Label &HJD &Phase &S/N &Label\\
    2450000+ &      &     &      &245000+ &&& \\ \hline
    July-August 1999 &&& &September 1999 &&& \\
    1383.7017 &0.7998 &248 &9  &1443.4928 &0.6915 &179 &4\\
    1384.7178 &0.9509 &238 &11 &1443.6184 &0.7102 &164 &5\\
    1385.7232 &0.1004 &217 &1  &1444.4788 &0.8382 &230 &6\\
    1386.7124 &0.2475 &229 &3  &1445.5656 &0.9998 &194 &7\\
    1387.7094 &0.3958 &200 &5  &1447.6084 &0.3036 &174 &1\\
    1388.7278 &0.5472 &181 &7  &1448.6212 &0.4542 &184 &2\\
    1389.7031 &0.6923 &178 &8  &1449.5979 &0.5994 &178 &3\\
    1390.7198 &0.8435 &151 &10 &&&&\\
    1391.7045 &0.9899 &243 &12 &&&&\\
    1392.7108 &0.1396 &258 &2  &&&&\\
    1393.7275 &0.2908 &189 &4  &&&&\\
    1394.7386 &0.4411 &206 &6  &&&&\\ \hline
    August 2001      &&& &August 2002 &&&\\
    2120.5503 &0.3796 &192 &3   &2507.6078 &0.9405 &139 &9 \\
    2121.5642 &0.5303 &225 &4   &2508.6582 &0.0967 &208 &1 \\
    2122.7102 &0.7008 &185 &6   &2509.5877 &0.2349 &200 &3 \\
    2123.6083 &0.8344 &196 &8   &2510.5909 &0.3842 &216 &4 \\
    2124.5644 &0.9765 &193 &10  &2511.5413 &0.5255 &195 &5 \\
    2125.6758 &0.1418 &177 &1   &2512.6316 &0.6876 &212 &7 \\
    2126.7146 &0.2963 &189 &2   &2513.5494 &0.8242 &179 &8 \\
    2128.6067 &0.5777 &166 &5   &2514.5854 &0.9782 &206 &10 \\
    2129.6470 &0.7324 &190 &7   &2515.5768 &0.1257 &207 &2 \\
    2130.6415 &0.8804 &208 &9   &2518.5749 &0.5715 &129 &6 \\ \hline
    November 2002    &&& &&&&\\
    2588.3964 &0.9549 &207 &7 &&&& \\
    2589.3764 &0.1007 &160 &1 &&&& \\
    2591.3567 &0.3951 &241 &3 &&&& \\
    2592.3705 &0.5459 &219 &4 &&&& \\
    2597.4333 &0.2988 &252 &2 &&&& \\
    2599.4798 &0.6031 &275 &5 &&&& \\
    2600.4896 &0.7532 &259 &6 &&&& \\
    \end{tabular}
  \end{center}
\end{table}

When a star rotates rapidly star spots modify the observed spectral
line profiles. As the star rotates, these bumps will move across the
absorption line profiles (Hackman et al. (2001); Kurster, M. (1993)). The surface imaging techique is basically to
trace these distortions and create a surface map of the star. We use
the surface imaging technique developed by Piskunov (the code {\sc
  INVERS7}, Piskunov et al. (1990); Piskunov (1991)).

We use the stellar model atmospheres of Kurucz (1993) for local line
profile calculations
for a set of temperatures and limb angles. This
table is then used for the disk integration of a given surface
temperature distribution. The surface imaging problem can be solved by
searching for a such surface temperature distribution that minimizes
the discrepancy function between the observations and the calculated
line profiles (Hackman et al. (2001)).

High resolution spectra of II Peg were measured in July-August 1999,
September 1999, August 2001, August 2002 and November 2002. All the observations were made using the
SOFIN high resolution echelle spectrograph at the 2.6
m Nordic Optical Telescope (NOT), La Palma, Spain.
The data
were acquired with the 2nd camera equipped with a CCD detector
of 1152$\times$298 pixels,
and the spectral region $6160-6210$\AA  was chosen for surface imaging. 
The observations are summarized in Table 1.

The spectral observations were reduced with the 4A software system
(Ilyin 2000). Bias, cosmic ray, flat field and scattered light
corrections, wavelength calibration and normalization were included in
the reduction process.

The stellar parameters used in the inversions were chosen
according to Berdyugina et al. (1998a), and read $T_{eff}$=4600 K,
$\xi_t=2.4$ kms$^{-1}$, $\zeta_t$=3.5 kms$^{-1}$, $P_{\rm
  orb}$=6.724333 d, $v \sin i$=22.6 kms$^{-1}$, $i=60^{\circ}$.

\begin{figure}
\begin{center}
\includegraphics[width=2.5in]{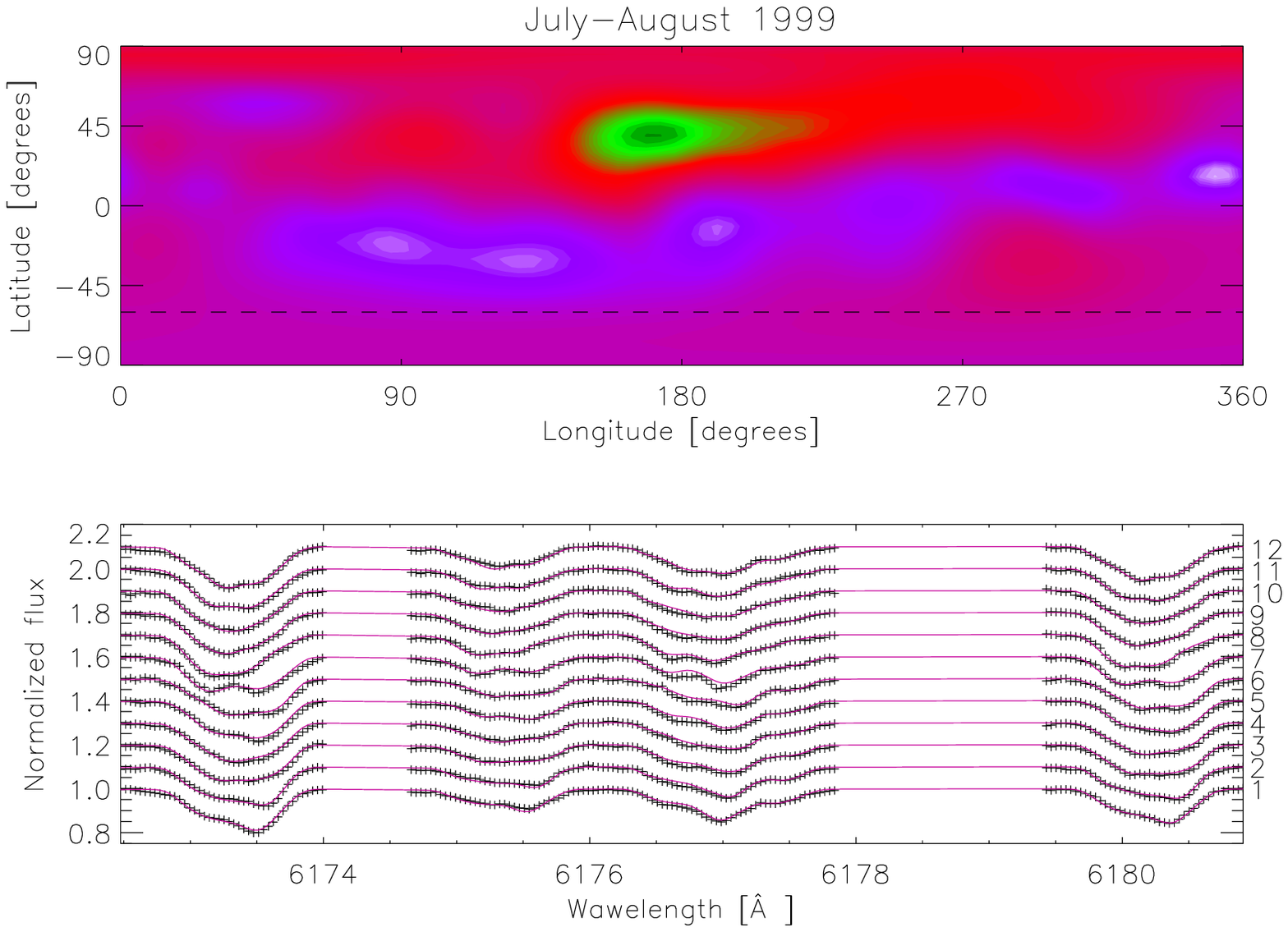}
\includegraphics[width=1.7in,angle=90]{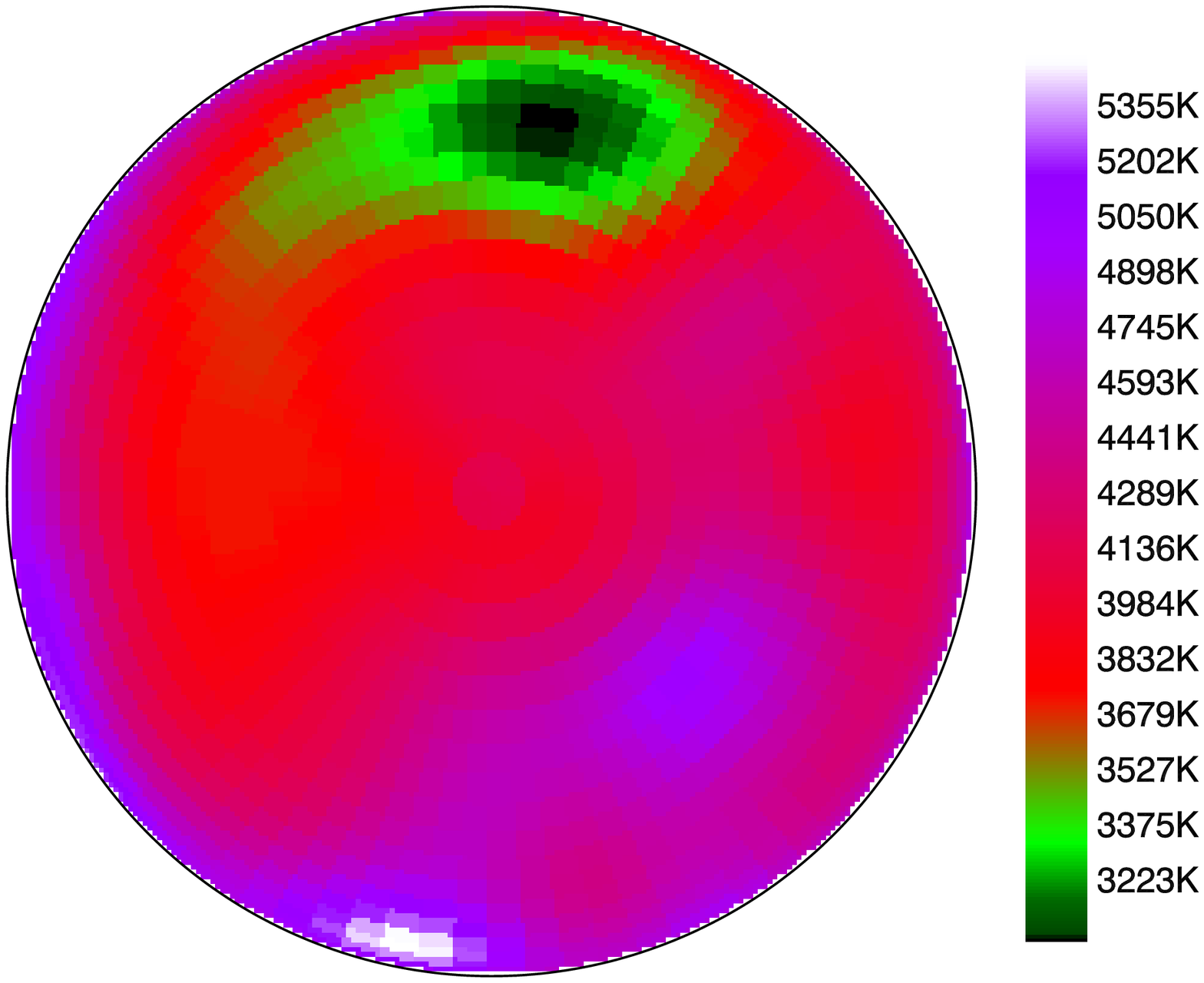}
\caption{Observing run during July - August 1999. Upper left panel:
  Mercator projection of the obtained surface temperature
  distribution. Lower left panel: observed (crosses) and calculated
  line profiles (solid lines). Panel on the right: Pole-on projection
  of the surface temperature distribution.}
\end{center}
\end{figure}

\begin{figure}
\begin{center}
\includegraphics[width=2.5in]{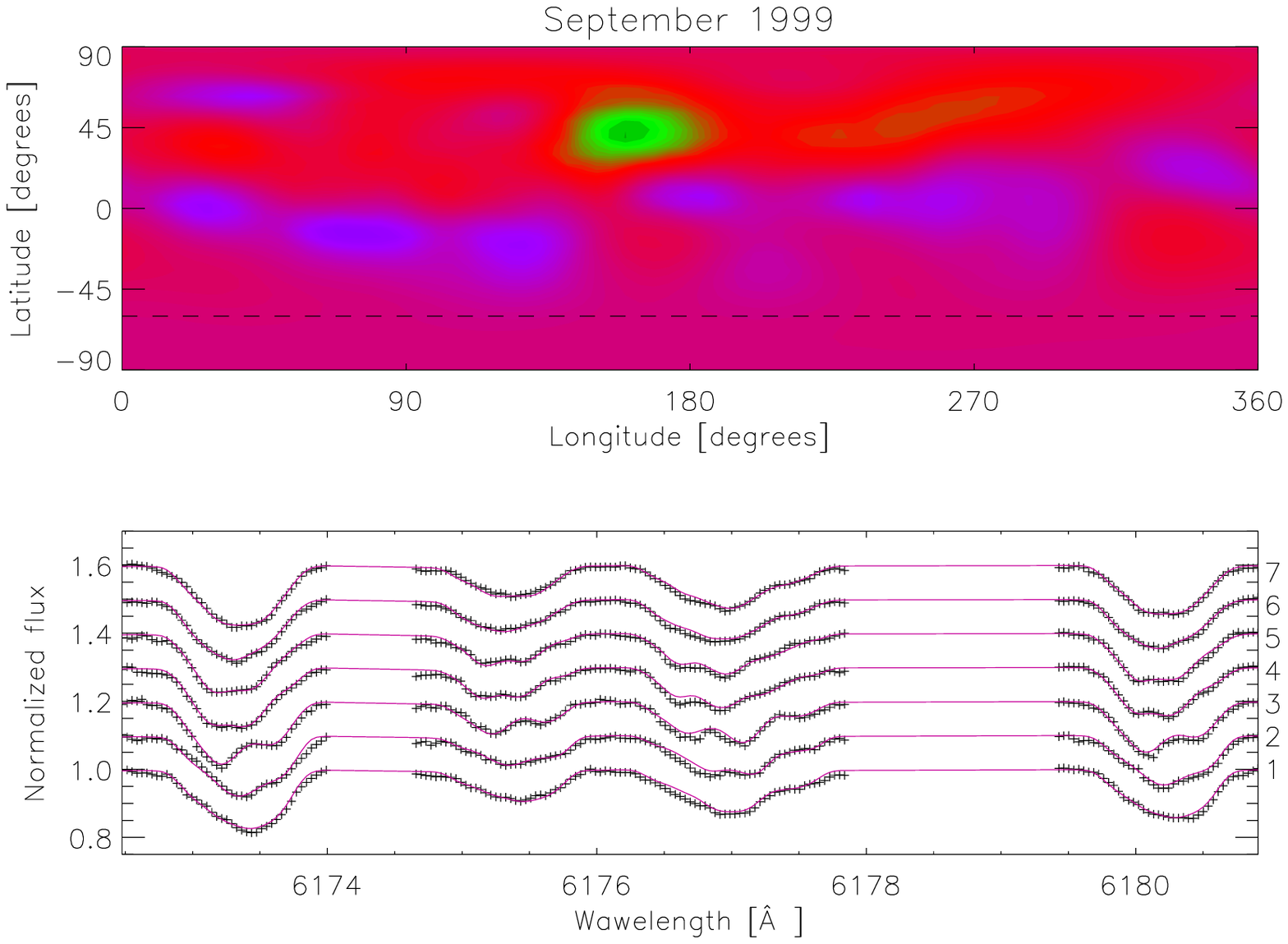}
\includegraphics[width=1.7in,angle=90]{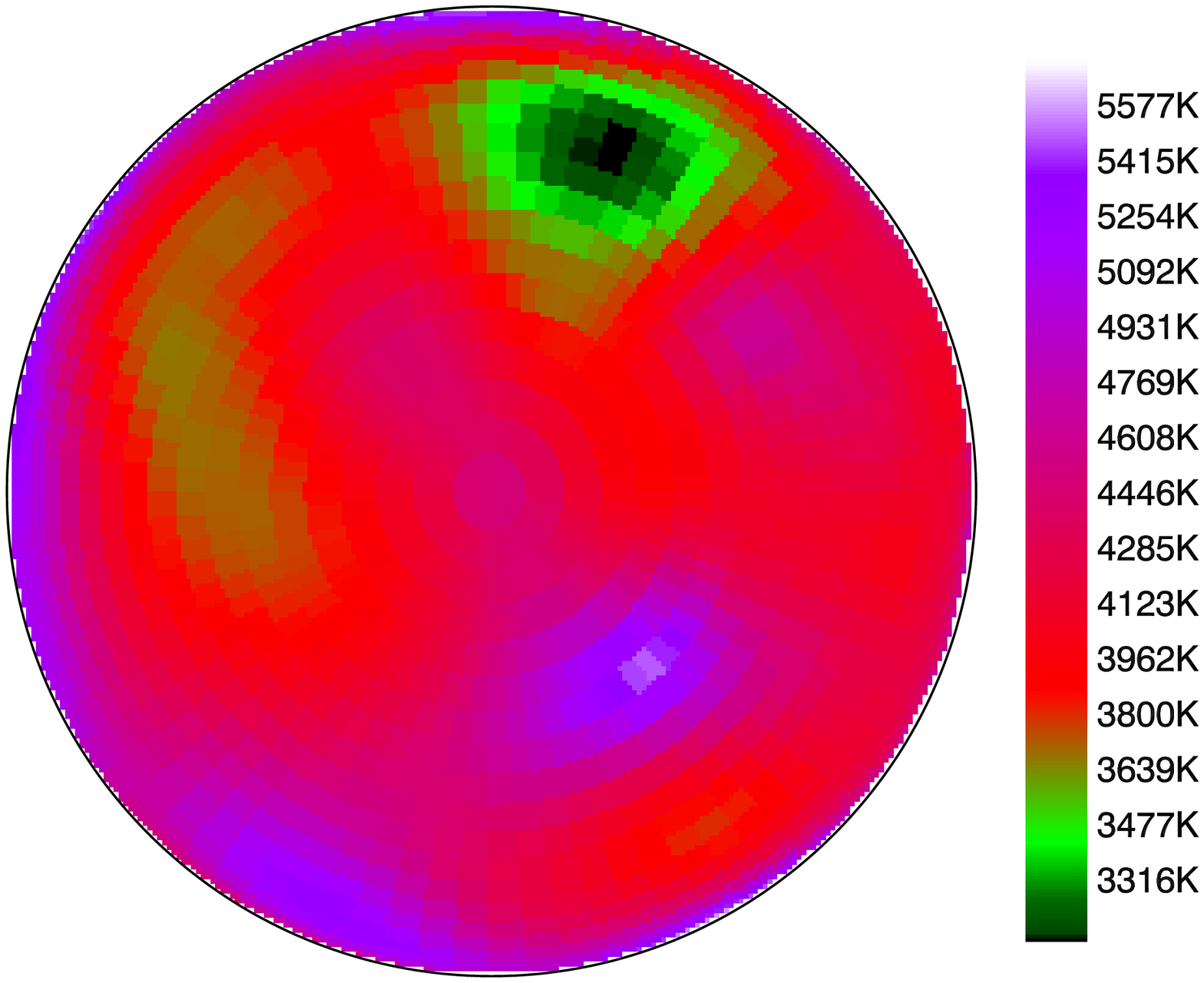}
\caption{Observing run during September 1999.}
\end{center}
\end{figure}

\begin{figure}
\begin{center}
\includegraphics[width=2.5in]{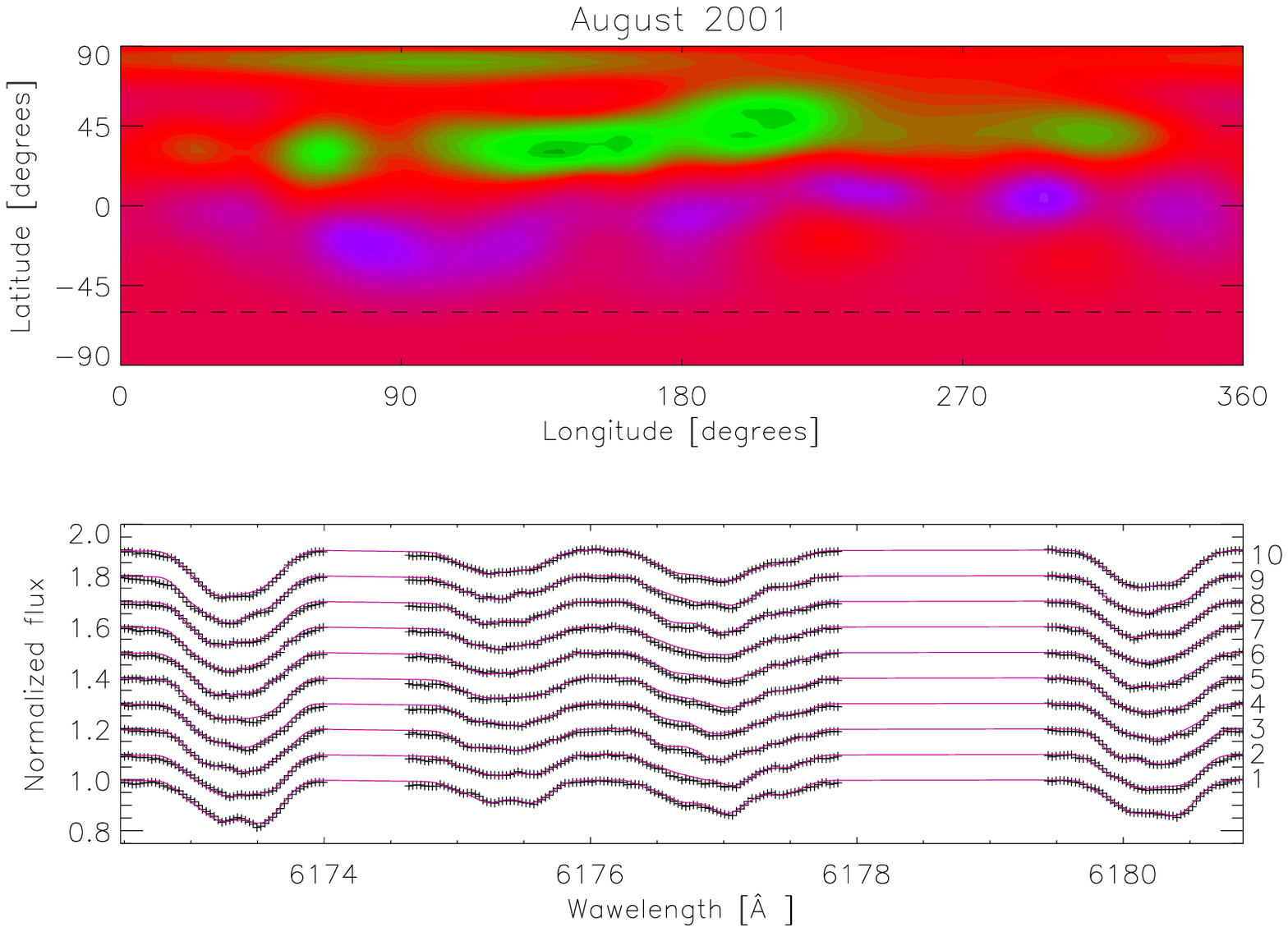}
\includegraphics[width=1.7in,angle=90]{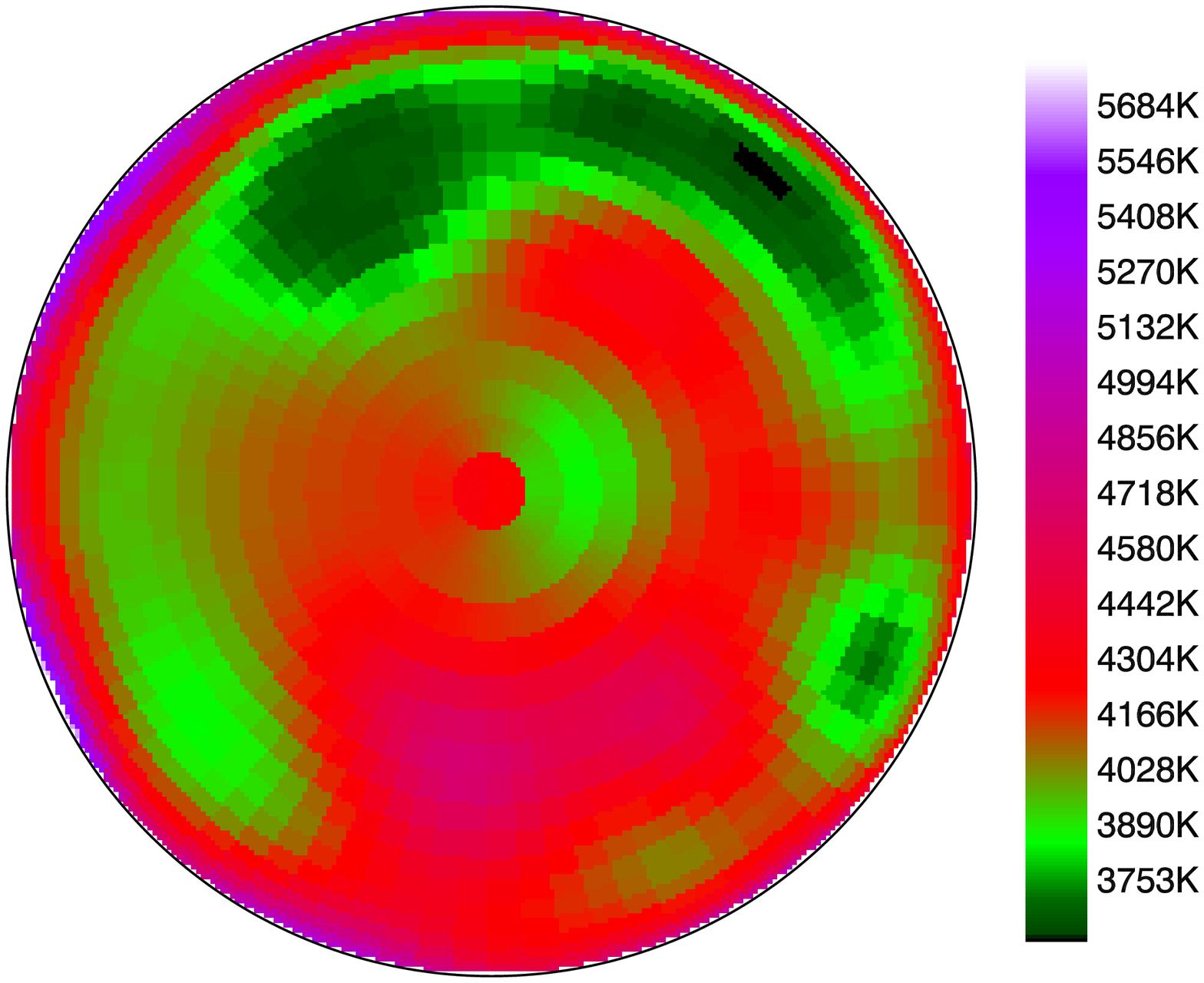}
\caption{Observing run during August 2001.}
\end{center}
\end{figure}

\begin{figure}
\begin{center}
\includegraphics[width=2.5in]{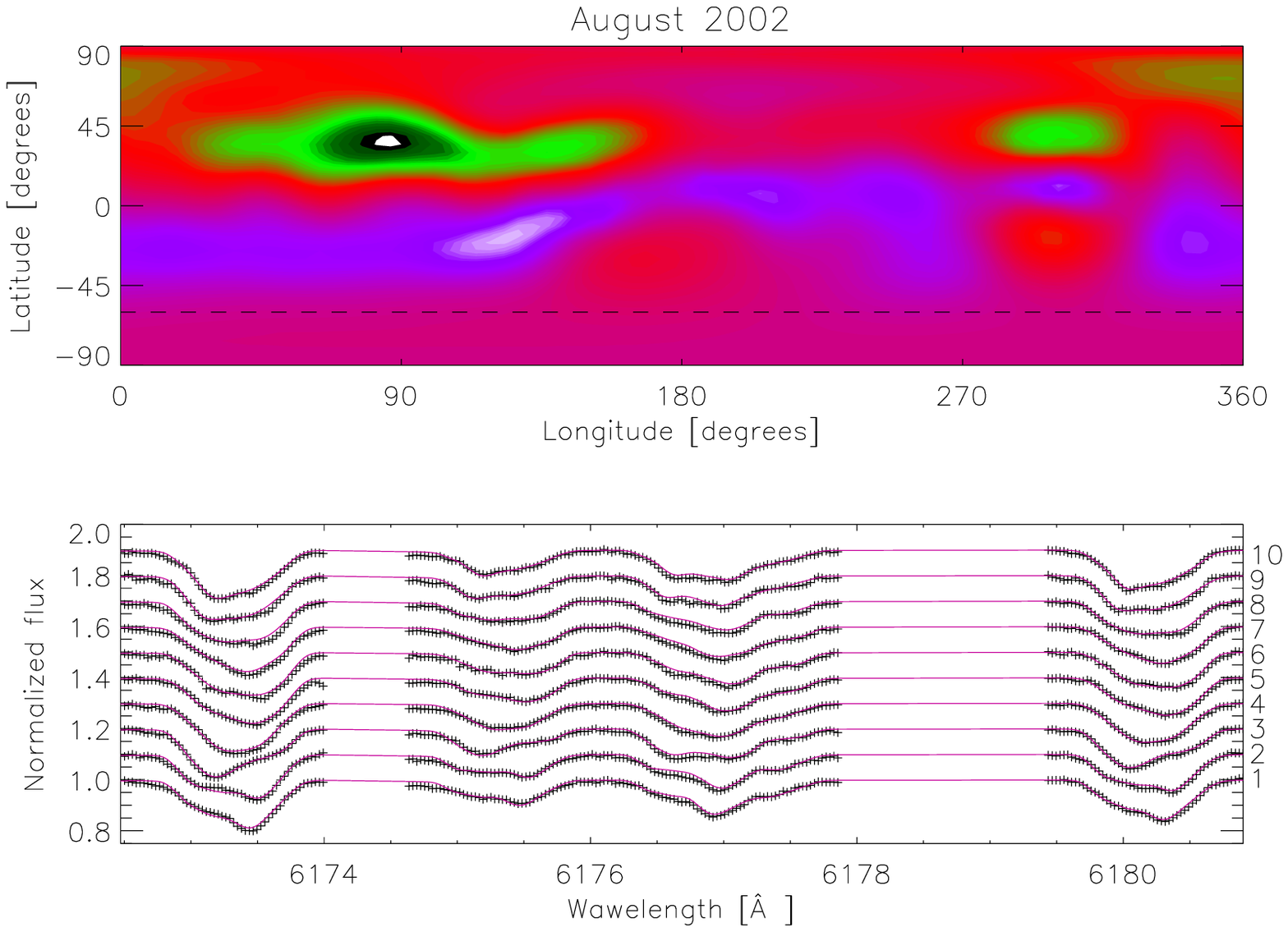}
\includegraphics[width=1.7in,angle=90]{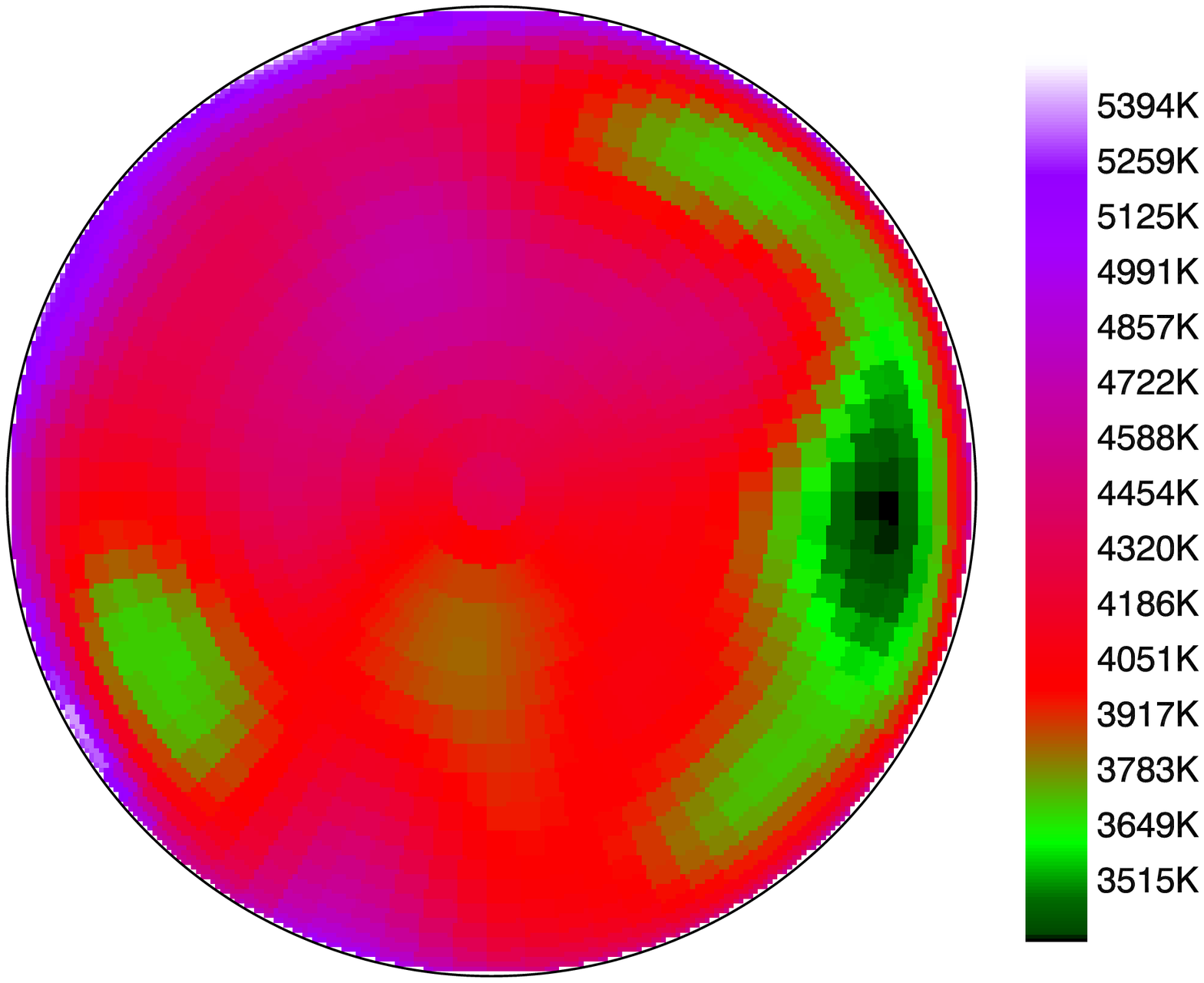}
\caption{Observing run during August 2002.}
\end{center}
\end{figure}

\begin{figure}
\begin{center}
\includegraphics[width=2.5in]{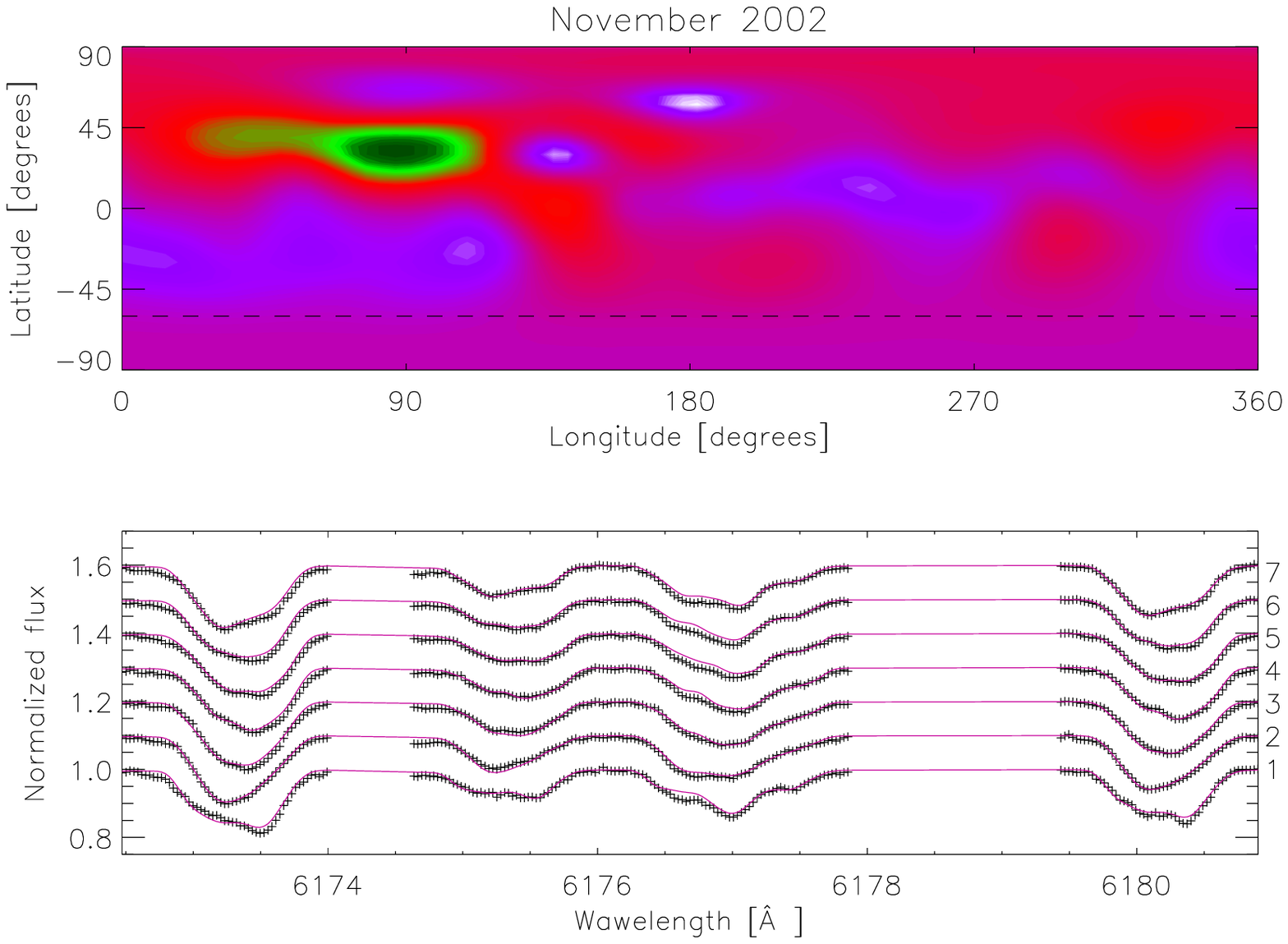}
\includegraphics[width=1.7in,angle=90]{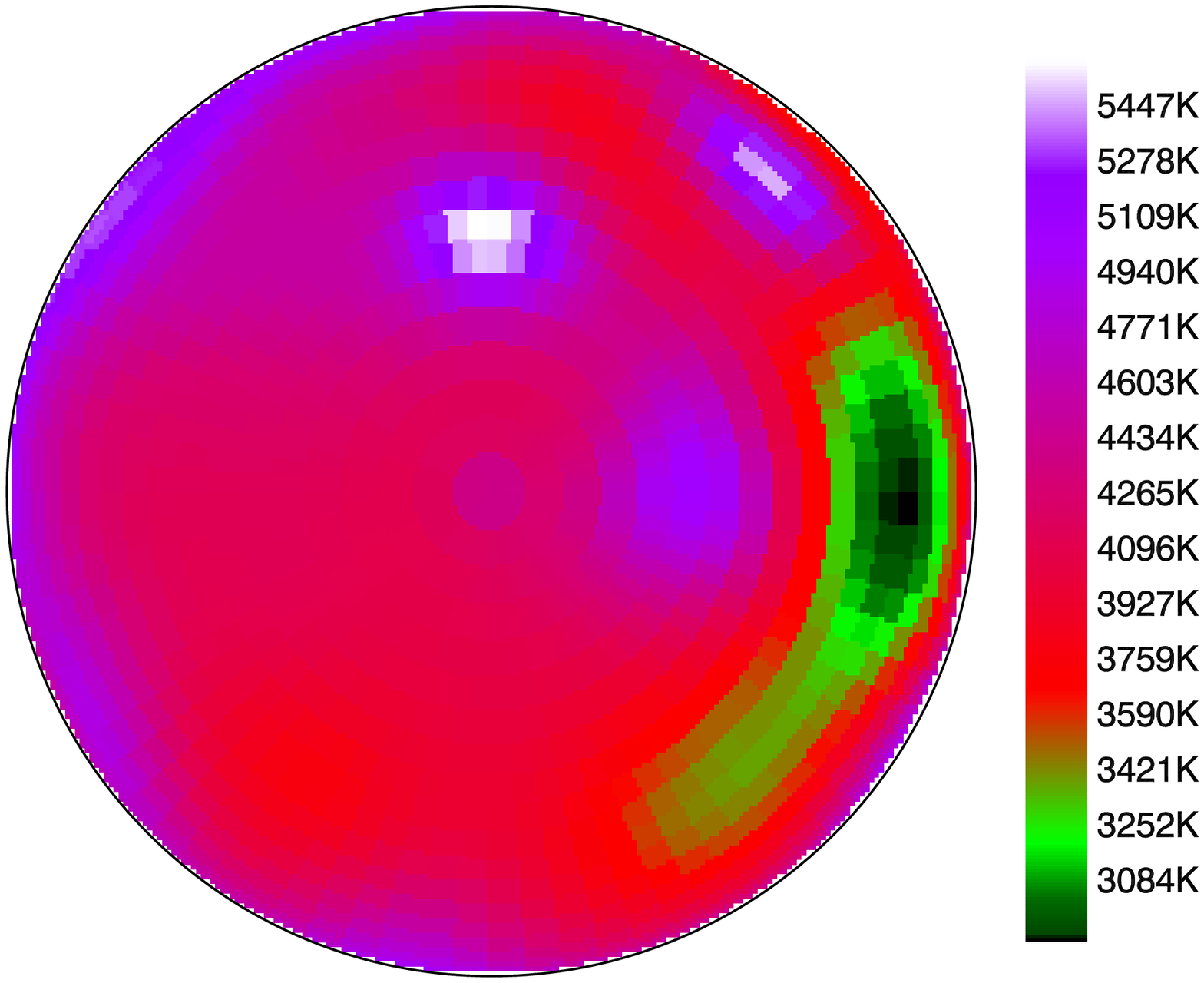}
\caption{Observing run during November 2002.}
\end{center}
\end{figure}

\begin{figure}
\begin{center}
\includegraphics{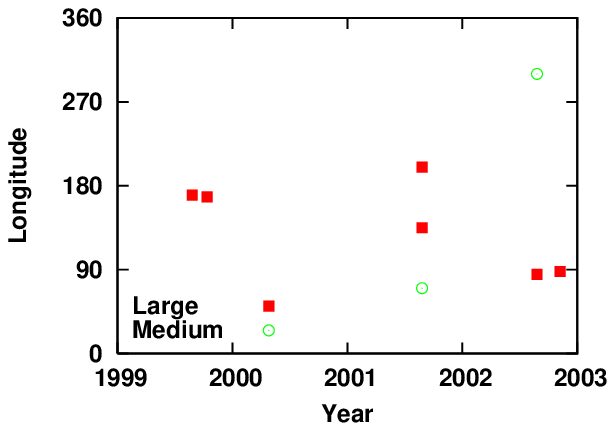}
\caption{The spot longitudes of II Peg versus time. Square symbols represent larger active regions and circles mark weak structures.}
\end{center}
\end{figure}

\section{Results and conclusions}

The following spectral lines were used in the surface temperature
inversions: Fe I 6173 \AA, Ni I 6175 \AA, Ni I 6177 \AA, and Fe I 6180
\AA. 
Spectral line parameters were obtained from the Vienna Atomic Line
Database (Kupka et al. 1999); the $log(gf)$ values of the two included Ni I
lines were modified from the standard value of $-0.53$ to $-0.2$ as
the standard values were found to produce much weaker absorption lines
than the actually observed ones. This is equivalent to an increase in the Ni abundancy, which is probably the real reason for the observed strong Ni lines. 

Figures 1-5 show the results of the inversions. During July-August
1999 (Fig. 1) only one spot is seen approximately at latitude
40$^{\circ}$. The minimum temperature inside the spot was measured to
occur at the approximate longitude of 170$^{\circ}$ or phase 0.47. Our
image is quite different from the one obtained by Gu et al. (2003) for
almost a simultaneous observing period, but a different spectral
region. Their inversions gave much larger, longitudinally extended,
spot structures around 170-290$^{\circ}$ at roughly 60$^{\circ}$ of
latitude.

In September 1999 our inversions also show only one strong spot that
has barely moved in the orbital reference frame (latitude 44$^{\circ}$
and longitude 160$^{\circ}$ or phase 0.44). There is a weaker,
longitudinally extended feature visible between 220-270$^{\circ}$ or
phase 0.61-0.75.

Almost two years later, in August 2001 (Fig. 3), the star exhibits
much more surface structures: three spots are visible in our image
(longitudes 70$^{\circ}$, 140$^{\circ}$, and 200$^{\circ}$ or phases
0.19, 0.39 and 0.56) at an approximate latitude of 40$^{\circ}$. The
inversions of Gu et al. (2003) for an observing run 5 months later
show much less structure, and the maximal activity seems to have moved
roughly to the other side of the star than what was seen in their
images during 1999 and 2000.

One year later, in August 2002 (Fig. 4) our maps show one strong spot
at the latitude 40$^{\circ}$ and longitude 80$^{\circ}$ or phase 0.22,
and a weaker one at 300$^{\circ}$ or phase 0.83. In November 2002
(Fig. 5) only the stronger spot is seen approximately at the same
location. Comparing the surface temperature distribution during the
observing seasons 1999 and 2002, the maximal spot activity seems to
have moved by roughly 100$^{\circ}$ in the orbital reference frame,
while very little drift of the spots can be seen during the
consequtive observing runs during 1999 and 2002. In between these two
'states' of only one strong spot at different location on the surface,
a much more complex distribution could be seen during August 2001. Our
images give some support to the conclusion of Gu et al. (2003) of a
significant change of the longitudinal spot distribution having
occured sometime during 2001 (Fig. 6). 

We plan to continue to study the spot
evolution on II Peg by analysing photometric and spectropolarimetric
observations of the object, both to invert the surface temperature
distribution, but also the magnetic field configuration of the object.

\acknowledgements{The authors acknowledge the hospitality of NORDITA
  during the programme 'Solar and stellar dynamos and cycles', and
  useful discussions with Dr. Oleg Kochukhov}

\end{document}